\documentclass[twocolumn,aps,pre,showpacs,floatfix,groupedaddress]{revtex4-1}
\usepackage{graphics}
\usepackage{graphicx}
\usepackage{amsmath}
\usepackage{amssymb}
\usepackage{ulem}
\usepackage{color}

\begin{document}
\title{Voronoi Cell Patterns: theoretical model and applications}

\author{Diego Luis Gonz\'alez}
\email[]{dgonzal2@umd.edu}
\affiliation{Department of Physics, University of Maryland, College Park, Maryland 20742-4111  USA}
\author{T. L. Einstein}
\email[]{einstein@umd.edu}
\affiliation{Department of Physics, University of Maryland, College Park, Maryland 20742-4111  USA}

\date{\today}


\begin{abstract}
We use a simple fragmentation model to describe the statistical behavior of the Voronoi cell patterns generated by a homogeneous and isotropic set of points in 1D and in 2D. In particular, we are interested in the distribution of sizes of these Voronoi cells. Our model is completely defined by two probability distributions in 1D and again in 2D, the probability to add a new point inside an existing cell and the probability that this new point is at a particular position relative to the preexisting point inside this cell.  In 1D the first distribution depends on a single parameter while the second distribution is defined through a fragmentation kernel; in 2D both distributions depend on a single parameter. The fragmentation kernel and the control parameters are closely related to the physical properties of the specific system under study. We use our model to describe the Voronoi cell patterns of several systems. Specifically, we study the island nucleation with irreversible attachment, the 1D car-parking problem, the formation of second-level administrative divisions, and the pattern formed by the Paris M\'etro stations.
\end{abstract}


\maketitle


\section{Introduction}\label{int}
\begin{figure}[htp]
\begin{center}
\includegraphics[scale=0.35]{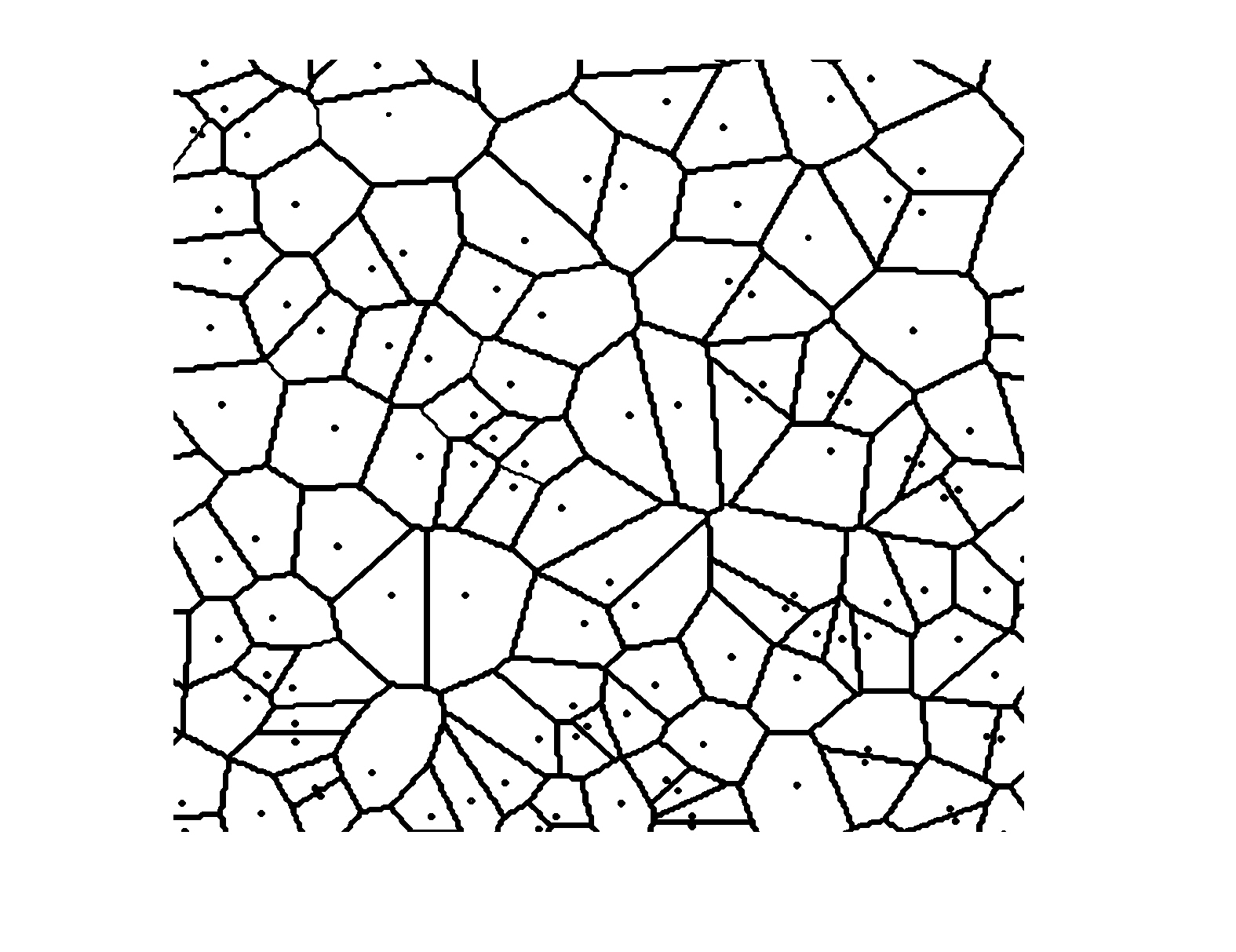}
\end{center}
\caption{Typical pattern of Poisson Voronoi cells. Note that in general the positions of the centers inside cells do not coincide with the geometrical centers of cells.}
\label{pattpv}
\end{figure}

Consider a set of points---usually called centers \cite{weaire,weaire1,mulheran} even though they are not geometric centers---on a discrete lattice. The Voronoi cell of a particular center $i$, is defined by all lattice points which are closer to $i$ than any other center. Figure~\ref{pattpv} shows a typical two-dimensional Voronoi pattern for the case where the positions of the centers are completely uncorrelated. This case is usually called Poisson Voronoi (PV). When the position of the centers are correlated, the system is called non-Poisson Voronoi (NPV).

Many different systems in nature resemble the PV patterns. Some examples can be found in areas such as ecology, astronomy, geology, biology, physics, and meteorology; see Refs.~\cite{weaire,weaire1,mercier,xue,mulheran,caer}. Applications of NPV patterns are not so extensive as those of the PV case. A couple of examples of NPV applications can be found in Refs.~\cite{mulheran,zaninetti,gonzalez7,blackman}.

One of the most important quantities in this system is the distribution of the sizes of the Voronoi cells $\hat{P}(S)$. The scaled size is defined as $s=S/\left\langle S\right\rangle$, where $\left\langle S\right\rangle$ is the average of $S$. The scaled size distribution is given by $P(s)=\left\langle S\right\rangle \hat{P}(s \left\langle S\right\rangle)$. There are many theoretical and numerical studies about PV systems \cite{kiang,weaire,weaire1,neda,hilhorst,hilhorst1,xua,gilbert}. In spite of this, the patterns formed by Voronoi cells have not been understood completely even in this simplest case, where the positions of the centers are not correlated. Most of our knowledge is based on empirical equations and numerical simulations. In fact, an analytical expression for $P(s)$ is known just for the 1D case with uncorrelated centers \cite{kiang,neda}.

In this paper we focus on NPV patterns in 1D and 2D. In Sec. \ref{secPV} we review some important properties of the PV cells. In Sec. \ref{secNPV} we propose a model to generate NPV cells for a homogeneous and isotropic set of points. In Sec. \ref{secAp} we provide several examples of different systems which can be described by NPV patterns. Finally, in Sec. \ref{secCon} we give conclusions.

\section{Poisson Voronoi cells}\label{secPV}
\subsection{One-dimensional Poisson Voronoi cells}
In one dimension, we have a ring divided in several sections called gaps. For the PV case, the positions of centers are completely random. Then, the probability density to find a gap with a length between $X$ and $X+dX$, $\hat{p}^{(0)}(X)$, is given by $\rho\,e^{-\rho\,X}$, where $\rho$ is the density of centers. The normalized gap size is defined as $x=X/\langle X \rangle$ where $\langle X \rangle=\rho^{-1}$ is the average gap size. Thus, the normalized gap size distribution can be written as $p^{(0)}(s)=e^{-x}$. By definition, $p^{(n)}(x)$ is the probability density to find a gap which starts and ends with a center subject to the condition that there are $n$ additional centers inside the gap. These distributions satisfy the normalization conditions $\int^{\infty}_0\,dx\,p^{(n)}(x)=1$ and $\int^{\infty}_0\,dx\,x\,p^{(n)}(x)=n+1$. Because sizes of the adjacent gaps are not correlated, it is possible to write the normalized spacing distributions for arbitrary values of $n$ in Laplace space as
\begin{equation}\label{gpn}
\tilde{p}^{(n)}(l)=\left(\tilde{p}^{(0)}(l)\right)^{n+1},
\end{equation}
where $\tilde{p}^{(n)}(l)=\int^{\infty}_0 dx\,e^{-x\,l}\,p^{(n)}(x)$ is the Laplace transform of $p^{(n)}(x)$~\cite{gonzalez}. Consequently, the normalized spacing distributions between centers are given by
\begin{equation}\label{p1}
p^{(n)}(x)=\frac{1}{n!}x^{n}e^{-x}.
\end{equation}
The pair correlation function has the simple form
\begin{equation}
g(x)=\sum^{\infty}_{n=0}p^{(n)}(x)=1.
\end{equation}
The size distribution of the Vononoi cells is related to the next-nearest-neighbor distribution according to \cite{gonzalez7}
\begin{equation}\label{CZ1d}
P(s)=2\,p^{(1)}(2s).
\end{equation}
This equation is a consequence of the one-dimensional nature of the ring. Therefore, the simple relation between $P(s)$ and $p^{(1)}(s)$ shown above is not valid for higher dimensions. Explicitly, the distribution of sizes of Voronoi cells in 1D is
\begin{equation}\label{h1}
P(s)=4\,s\,e^{-2\,s}.
\end{equation}

\subsection{Two-dimensional Poisson Voronoi cells}
Let $p^{(n)}(R)$ be the radial probability density that, given an island at $r=0$, its $(n+1)^{th}$ neighbor is between $R$ and $R+dR$. Given that there are $n$ additional centers inside of the circle of radius $R$, the radial spacing distribution, $p^{(n)}(R)$, can be calculated as follows. As usual, $\rho$ is the density of centers. On average, the total number of centers, $c(R)$, within a disk with radius $R$ is $2\pi \rho\,R$. It is well known \cite{redner} that $c(R)$ and $\hat{p}^{(0)}(R)$ are related by
\begin{equation}\label{pcr}
\hat{p}^{(0)}(R)=c(R)e^{-\int^{R}_{0}dr\, 2\pi r\, c(r)}.
\end {equation}
More details about this equation are given in the Appendix. Since $c(R)=2\,\pi\,R\,\rho$, $\hat{p}^{(0)}(R)$ has the simple form
\begin{equation}
\hat{p}^{(0)}(R)=2\,\pi\,\rho\,R\,e^{-\pi\,\rho\,R^2}.
\end {equation}
The next radial distribution $\hat{p}^{(1)}(R)$ is given by
\begin{equation}\label{pr1}
\hat{p}^{(1)}(R)= \,2\,\pi\,\rho\,R \int_0^{R}dR^{\prime}\,\hat{p}^{(0)}(R^{\prime})\hat{Q}(R^{\prime},R) ,
\end {equation}
where $\hat{Q}(R^{\prime},R)=e^{-\pi\,\rho(R^2-{R^{\prime}}^2)}$ is the probability density to have the annulus $R^{\prime}\leq r \leq R$ free of centers. The integral in Eq.~(\ref{pr1}) can be calculated straightforwardly to give
\begin{equation}
\hat{p}^{(1)}(R)=2\,\pi^2\,\rho^2\,R^3e^{-\pi\,\rho\,R^2}.
\end {equation}
Following the previous procedure, one can show that, for arbitrary $n$,
\begin{equation}\label{pr2d}
\hat{p}^{(n)}(R)=2\frac{(\pi\,\rho)^{n+1}}{n!} R^{2n+1}e^{-\pi\,\rho\,R^2}.
\end{equation}

As for multineighbor spacing distributions in 1D \cite{hansen,gonzalez,gonzalez4}, the radial distributions have information about the structure of the system. For example, since
\begin{equation}\label{gr2d}
2\,\pi\,R\,\rho\,\hat{g}(R)=\sum^{\infty}_{n=0}\hat{p}^{(n)}(R),
\end{equation}
it is possible to calculate the radial distribution function $\hat{g}(R)$. From Eqs.~(\ref{pr2d}) and (\ref{gr2d}) it is easy to find
\begin{equation}
\hat{g}(R)=e^{-\pi\,\rho\,R^2}\sum^{\infty}_{n=0}\frac{(\pi\,\rho\,R^2)^{n}}{n!}=1.
\end{equation}
This result is consistent with this case of uncorrelated centers $g(R)=1$. In general the concentration of centers $c(R)$ can be extracted from $\hat{p}^{(0)}(R)$. From Eq.~(\ref{pcr}) it follows
\begin{equation}
c(R)=\frac{\hat{p}^{(0)}(R)}{\int^{\infty}_R dr \hat{p}^{(0)}(r)}.
\end{equation}

Even though we know the exact expression for the radial spacing distributions for arbitrary values of $n$ in the PV case, the exact functional form of $P(s)$ is not known for $d\geq2$. In fact, just a few exact analytical results are known in this case. One of them was reported in 1962 by Gilbert \cite{gilbert}, who showed that the second moment of $P(S)$ is $0.280 \langle S \rangle$.

\begin{figure}[htp]
\begin{center}
\includegraphics[scale=0.30]{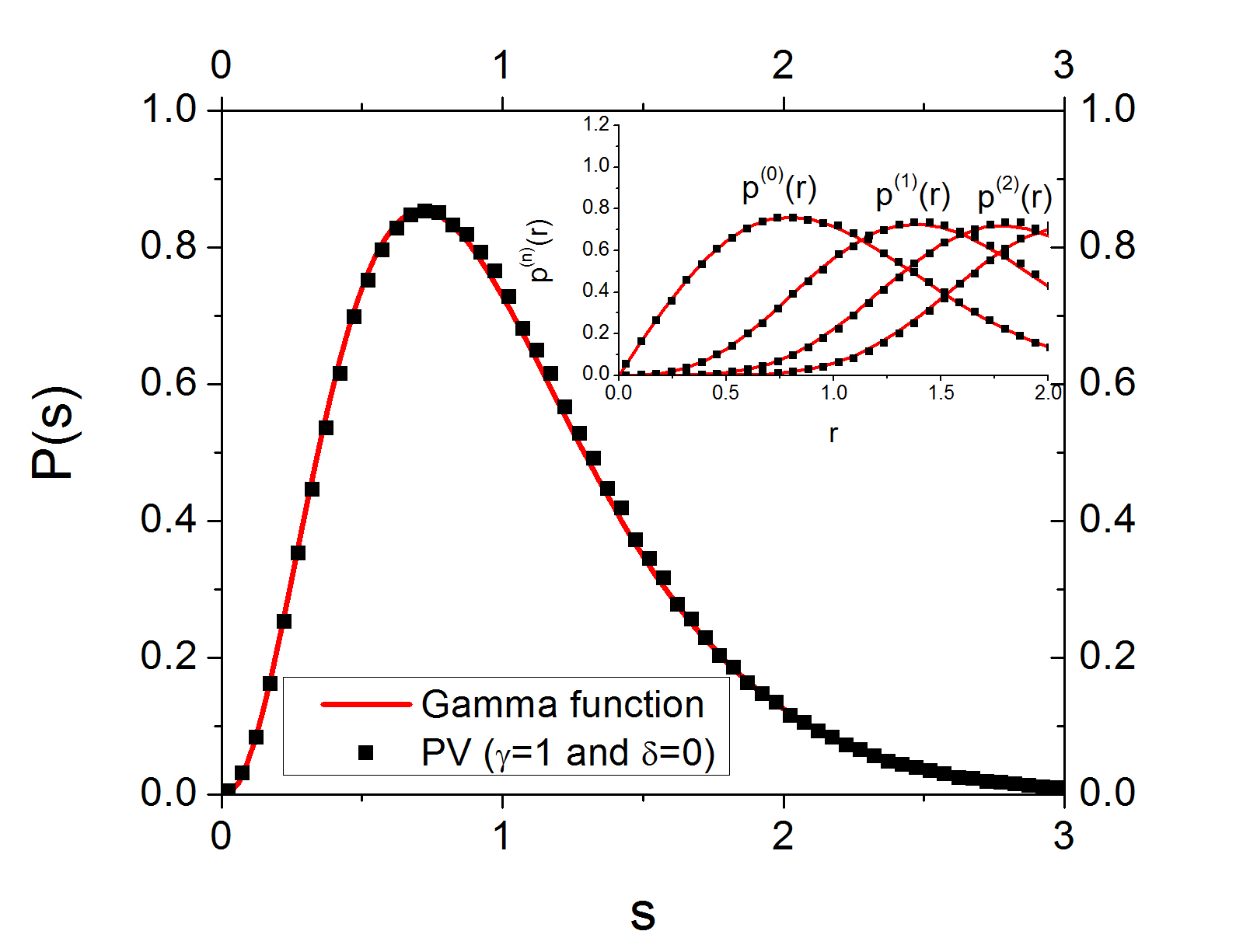}
\end{center}
\caption{(Color online) Poisson Voronoi cell-size distribution $P(s)$ with $d=2$. The agreement between the numerical results and Eq.~(\ref{ps2d}) is excellent. The inset shows the radial distribution functions $p^{(n)}(r)$. Red lines correspond to Eqs.~(\ref{pr2d}) and (\ref{ps2d}).}
\label{pvoronoi2d}
\end{figure}

As mentioned previously, obtaining the exact expression of $P(s)$ for $d>1$ is quite complicated partially due geometrical complications. In $d=1$ each new center divides just one of the existing Voronoi cells to form two new cells. This fact leads to the simple relation between $P(x)$ and $p^{(1)}(x)$ given by Eq.~(\ref{CZ1d}). In higher dimensions this does not apply; the Voronoi cell of a new center is formed at the expense of several preexisting Voronoi cells. An analogous relation to Eq.~(\ref{CZ1d}) for $d>1$ remains unknown, and it could involve several $p^{(n)}(r)$ in a non-trivial way. However, it is well accepted that $P(s)$ can be approximated by the gamma distribution $\Pi_{\alpha}(s)$. Based on extensive numerical simulations for $d=1,2,3$, Ferenc and N\'eda \cite{neda} proposed
\begin{equation}\label{ps2d}
P(s)\approx \Pi_{\alpha}(s)=\frac{\alpha^{\alpha}}{\Gamma(\alpha)}s^{\alpha-1}e^{-\alpha \, s}
\end{equation}
where $\alpha=(3d\! +\! 1)/2$. Note that for $d=1$ we recover Eq.~(\ref{h1}). The agreement between Eq.~(\ref{ps2d}) and the numerical results for $P(s)$ is excellent, as seen in Fig.~\ref{pvoronoi2d}.

\section{Non-Poissonian Voronoi cells}\label{secNPV}
More complicate behavior arises when the centers are correlated in some way. Although there are few studies \cite{zaninetti,mulheran,gonzalez7,blackman} about such systems, the NPV patterns can be used to describe qualitatively and quantitative many different systems, as we shall see.

\subsection{One-dimensional non-Poissonian Voronoi cells}
To generate a one-dimensional NPV set of centers we proceed as follows. Let $p^{g}(x)$ be the probability density to put a new center inside a gap with a scaled size $x$. In a similar way, $p^{x}(x)$ is the probability density that the new center is placed at the position $x$ with respect to the center at the left of the gap. This kind of model proved fruitful in studying the spatial structure of the one-dimensional point-island model for epitaxial growth. In fact,  a suitable choice of $p^{g}(x)$ and $p^{x}(x)$ leads to an excellent description of the physical properties of this system \cite{gonzalez7,blackman}. The gap size distribution, $p^{(0)}(x)$, was shown there to satisfy the equation
\begin{equation}\label{ed1d}
x \frac{dp^{(0)}(x)}{dx}+2\,p^{(0)}(x)=-p^{g}(x)+2\,p^{x}(x),
\end{equation}
In particular, the probability to put a new center inside a gap has the form
\begin{equation}\label{pa}
p^{g}(x)= \frac{x^{\gamma}}{\mu_{\gamma}} p^{(0)}(x),
\end{equation}
with $\mu_{\gamma}$ the $\gamma^{th}$ moment of $p^{(0)}(x)$. The probability density $p^{x}(x)$ can be written as
\begin{equation}\label{pc}
p^{x}(x)= \int^{\infty}_{x}dz\, \frac{1}{z}p^{g}(z)p^{\Lambda}(x/z),
\end{equation}
where $p^{\Lambda}(x/z)dx/z$ is the conditional probability that, given a particular gap with size $z$, the new center is placed inside $[x,x+dx]$.
Unfortunately, in most cases the integro-differential given by Eqs.~(\ref{ed1d}), (\ref{pa}) and (\ref{pc}) cannot be solved analytically. However, this system can be simulated numerically without major difficulties \cite{sim1}.

Nonetheless, some exact results can be extracted from Eq.~(\ref{ed1d}). If we choose $\gamma=1$ and $p^{\Lambda}(\lambda=x/z)=1$, the probability to put a new center onto an empty site is the same for all empty sites on the lattice. Then, with this selection of $\gamma$ and $p^{\Lambda}(\lambda)$, we recover the one-dimensional PV case discussed previously in Sec. \ref{secPV}-A. As mentioned, the position of the centers in the PV case are totally uncorrelated, and $p^{(n)}(x)$ is given by Eq.~(\ref{p1}), while $P(s)$ is given by Eq.~(\ref{h1}).

In general, for $\gamma\geq1$ and $p^{\Lambda}(x/z)=1$, the solution of Eq.~(\ref{ed1d}) can be calculated as follows. The simple form of the kernel $p^{\Lambda}(\lambda)$ leads to $dp^{x}(x)/dx=-x^{\gamma-1} p^{(0)}(x)/\mu_{\gamma}$. Then, differentiating Eq.~(\ref{ed1d}) we have
\begin{equation}
x \frac{d^2p^{(0)}(x)}{dx^2}+\left(3+\frac{x^{\gamma}}{\mu_{\gamma}} \right)\frac{dp^{x}(x)}{dx}+\left(2+\gamma\right)\frac{x^{\gamma-1}}{\mu_{\gamma}} p^{(0)}(x)=0.
\end{equation}
After some algebra the above equation can be written as
\begin{equation}
\left(\frac{d}{dx}+\frac{2}{x}\right)\left(x \frac{dp^{(0)}(x)}{dx}+\frac{x^{\gamma}}{\mu_{\gamma}} p^{(0)}(x)\right)=0,
\end{equation}
whose general solution is
\begin{equation}\label{eqper}
p^{(0)}(x)=\frac{1}{\Gamma\left(1+\frac{1}{\gamma}\right)\left(\mu_{\gamma}\,\gamma\right)^{\frac{1}{\gamma}}} \,e^{-\frac{x^{\gamma}}{\gamma\,\mu_{\gamma}}},
\end{equation}
with $\mu_{\gamma}=\left[\Gamma(1/\gamma)/\Gamma(2/\gamma)\right]^{\gamma}/\gamma$. In this case $p^{x}(x)$ is also given by Eq.~(\ref{eqper}). Note that $p^{(0)}(0)\neq0$. For this case higher spacing distributions, in particular $P(s)$, cannot be calculated easily.

Consider now a more general case where $p^{\Lambda}(x/z)$ depends on $x$ and $z$. We restrict our work to functions which are symmetrical about $\lambda=x/z=1/2$. This symmetry property comes from the fact that in the absence of an external drift (e.g., a field), $p^{\Lambda}(1-x/z)$ must be equal to $p^{\Lambda}(x/z)$. Furthermore, we impose the additional condition $p^{\Lambda}(0)=p^{\Lambda}(1)=0$. This property implies that the probability to place a new center near an existing one is small.

For large values of $x$, $dp^{(0)}(x)/dx$ is negative. Then, in this regime the behavior of $p^{(0)}(x)$ is dominated by $p^{g}(x)$. Consequently, for $x\gg1$, Eq.~(\ref{ed1d}) takes the form
\begin{equation}
x \frac{dp^{(0)}(x)}{dx}\approx-\frac{x^{\gamma}}{\mu_{\gamma}} p^{(0)}(x),
\end{equation}
which implies $p^{(0)}(x)\propto \mathrm{exp}\left({-x^{\gamma}/(\gamma\,\mu_{\gamma})}\right)$ and $p^{(0)}(x)\gg p^{x}(x)$. A first correction to this formula can be obtained by using the ansatz $p^{(0)}(x)\propto f(x)\,\mathrm{exp}\left({-x^{\gamma}/(\gamma\,\mu_{\gamma})}\right)$ in Eq.~(\ref{ed1d}). This procedure gives the differential equation
\begin{equation}
x\,\frac{df(x)}{dx}+2\,f(x)=0.
\end{equation}
We conclude that $p^{(0)}(x)\propto x^{-2} \mathrm{exp}\left({-x^{\gamma}/(\gamma\,\mu_{\gamma})}\right)$. Then the behavior of  $p^{(0)}(x)$ for large values of $x$ is completely determined by the parameter $\gamma$. Furthermore, in the limit $x\gg1$ the solution of Eq.~(\ref{ed1d}) does not depend on $p^{\Lambda}(\lambda)$. However, $p^{\Lambda}(\lambda)$ controls the behavior of $p^{(0)}(x)$ for small values of $x$. In general, for the kind of functions considered here, for $\lambda\ll1$ we have $p^{\Lambda}(\lambda)\sim \lambda^\zeta$, with $\zeta$ a constant which depends on the functional form of $p^{\Lambda}(\lambda)$. A series expansion of Eqs.~(\ref{ed1d}) and (\ref{pc}) shows that $p^{(0)}(x)\sim x^\zeta$ and $p^{x}(x)\sim x^\zeta$. It is clear that the vanishing condition imposed on $p^{\Lambda}(\lambda)$ for $\lambda=0$ leads to $p^{(0)}(0)=0$. The effective entropic ``repulsion force'' between centers is determined by the parameter $\zeta$ given by the series expansion of $p^{\Lambda}(\lambda)$ around $\lambda=0$.

\subsection{Two-dimensional non-Poissonian Voronoi cells}
In the case $d=2$ we proceed as follows. Let $q^{c}(s)$ be the probability density to put a new center within a Voronoi cell having a scaled area $s$. Explicitly, we consider the general form
\begin{equation}\label{pgamma}
q^{c}(s)=\frac{s^{\gamma}}{\tilde{\mu}_{\gamma}} P(s),
\end{equation}
where $\tilde{\mu}_{\gamma}$ is the $\gamma^{th}$ moment of $P(s)$. In a similar way, we define $q^{r}({\rm {\bf r}},s)$ as the probability density that, for a particular cell with scaled size $s$, the new center is located at a position {\bf r} with respect to the center of the preexisting cell. For the sake of simplicity, we consider just the isotropic case where (with $r \equiv |{\rm {\bf r}}|$)
\begin{equation}\label{palfa}
q^{r}({\rm {\bf r}},s)\sim r^{\delta}.
\end{equation}
In this simplification, the probability to put a new center inside a cell depends on the cell itself, regardless of the positions of neighboring centers or the areas of their surrounding cells. The functional form of $q^{r}({\rm {\bf r}},s)$ depends on the shape of the Voronoi cell. For example, in the case of a circular Voronoi cell with scaled area $s$, we have
\begin{equation}
q^{r}({\rm {\bf r}},s)=\frac{\delta+2}{2\,s^{1+\frac{\delta}{2}}}\pi^{\frac{\delta}{2}}r^{\delta}.
\end{equation}
The simplest case is $\gamma=1$ and $\delta=0$, for which $q^{r}({\rm {\bf r}},s)\propto1/s$. In this case every empty point of the lattice has the same probability to receive a new center. This, of course, corresponds to the PV case discussed in Sec.~\ref{secPV}-B and shown in Fig.~\ref{pvoronoi2d}.

From a Taylor expansion of Eq.~(\ref{pcr}) around $R=0$, it is clear that the behavior of $\hat{p}^{(0)}(R)$ is controlled by the functional form of $c(R)$. For small values of $R$, it is reasonable to suppose that the concentration of particles within a distance $R$ and $R+dR$ from a given center is proportional to the product of $q^{r}({\rm {\bf R}},s)$ and the area $dA=2\,\pi\,R\,dR$; then 
\begin{equation}
c(R)\,dR\sim q^{r}({\rm {\bf R}},s)\,dA\sim R^{\delta+1}dR.
\end{equation}
Then, $\hat{p}^{(0)}(R)\sim R^{\delta+1}$, and $\delta$ controls the effective ``repulsion force'' between centers. Using this simple result in Eq.~(\ref{gr2d}) we conclude that, for $R\ll1$, $\hat{g}(R)\sim R^{\delta}$. On the other hand, the behavior of $\hat{p}^{(0)}(R)$ for large values of $R$ depends on the behavior of the integral $\int^{R}_0 d\xi\,c(\xi)$.

The equivalent of Eq.~(\ref{ed1d}) for the 2D case cannot be written easily in terms of known quantities. Following Ref.~\cite{evans5}, the effect of a new center on $P(s)$ can be written as
\begin{equation}\label{psevol2d}
s\,\frac{dP(s)}{ds}+2\,P(s)=M\,p^{+}(s)-M\,p^{A}(s)+p^{*}(s),
\end{equation}
where $M$ is the average number of preexisting Voronoi cells overlapped by the Voronoi cell generated by the new center; $p^{+}(s)$ is the probability density that the new center reduces the Voronoi cell size of a preexisting cell to $s$, $p^{A}(s)$ is the probability density that the Voronoi cell of the new center overlaps a preexisting cell with size $s$; and $p^{*}(s)$ is the probability density that the new Voronoi cell has size $s$.

\begin{figure}[b]
\begin{center}
$\begin{array}{cc}
\includegraphics[scale=0.15]{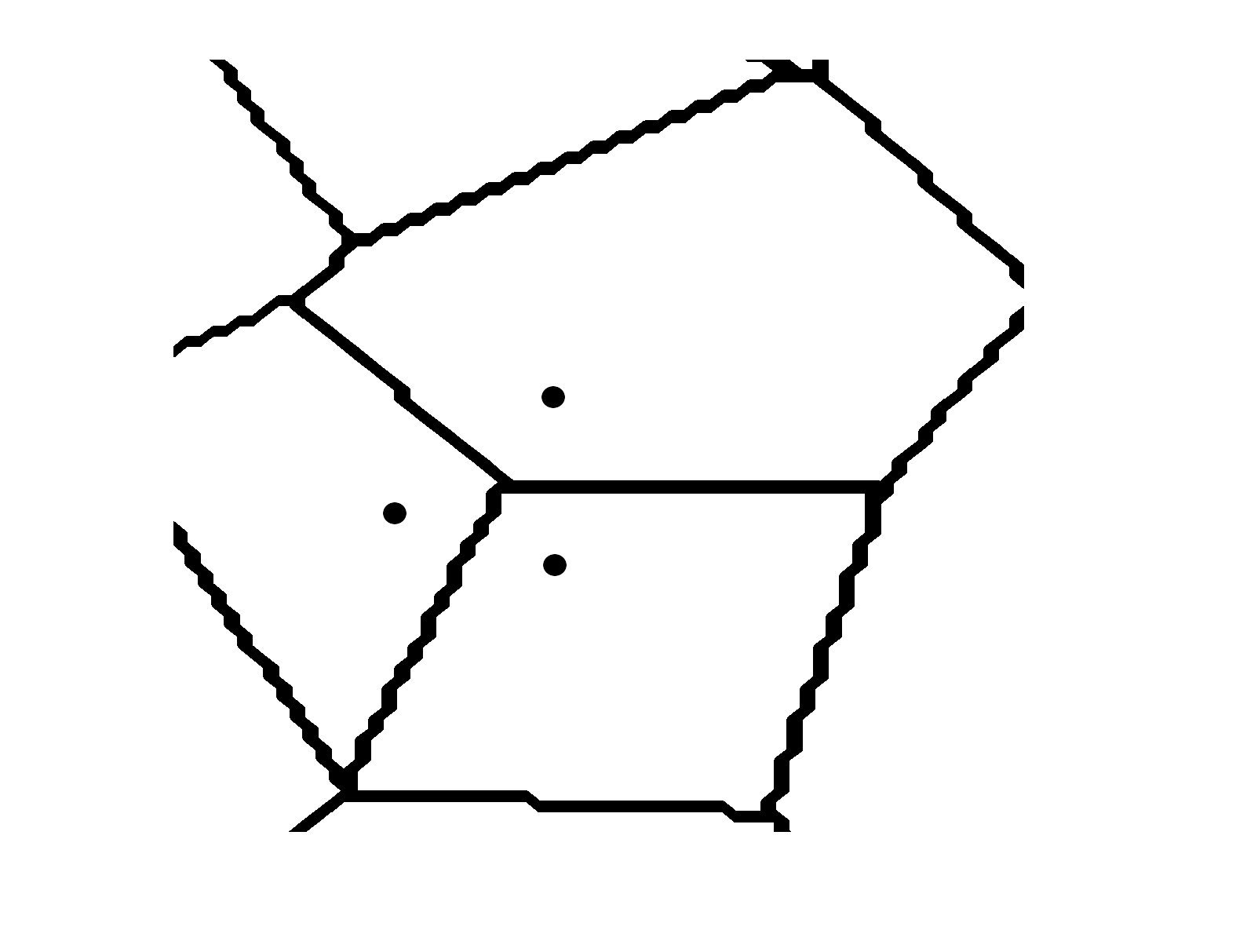}&
\includegraphics[scale=0.15]{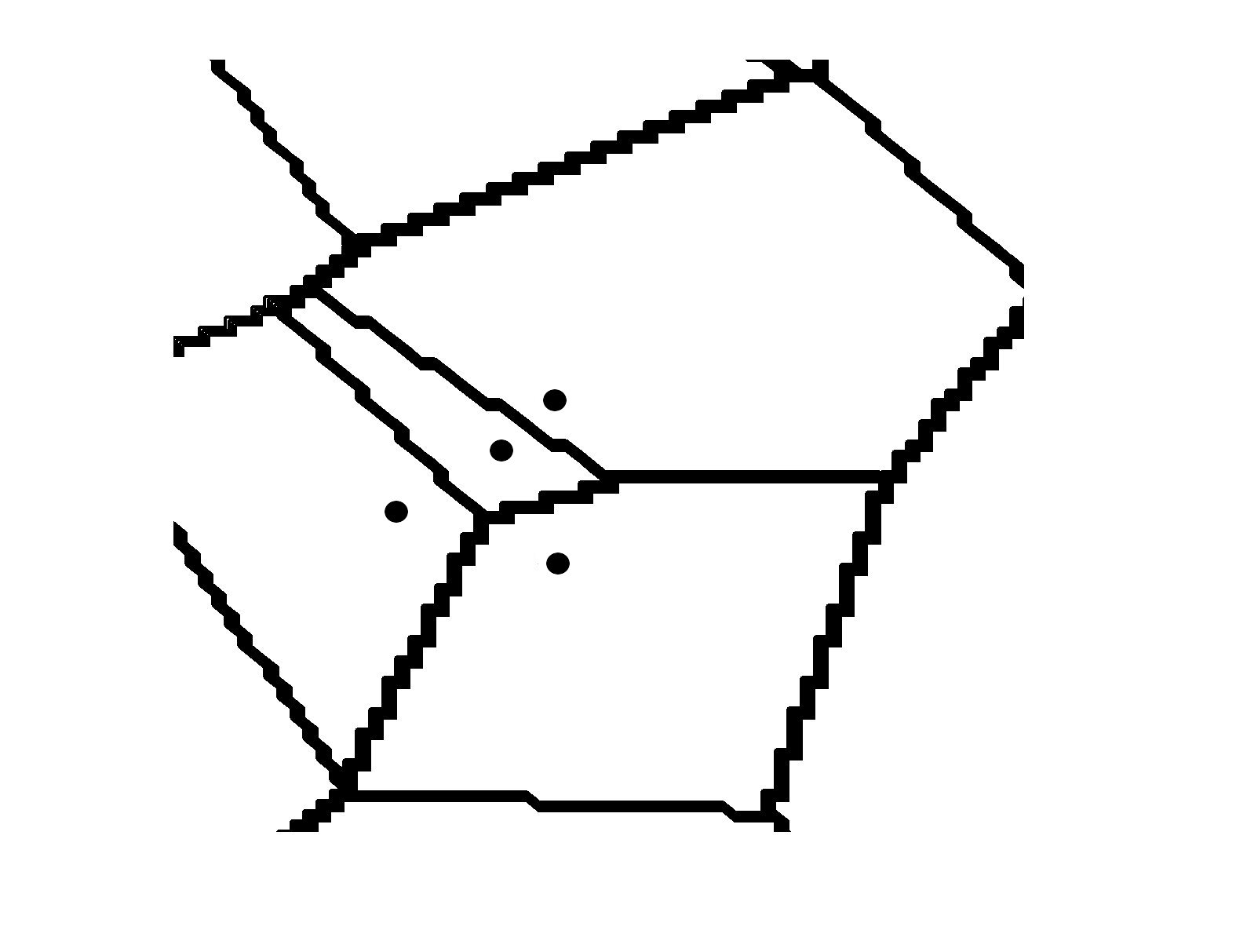}\\
(a) & (b) \\
\end{array}$
\end{center}
\caption{In (a) is the initial configuration of centers, while in (b) the effect of a new center on the preexisting Voronoi cells is shown. Note that in this particular case, the new Voronoi cell is almost completely defined by the first three neighbors of the new center.}
\label{cpzab}
\end{figure}

From their definitions, $q^{c}(s)$ and $p^{A}(s)$ are clearly related: $q^{c}(s)$ takes into account the destruction of a Voronoi cell by the direct impact of a new center on that cell, while $p^{A}(s)$  is more general and expresses that the Voronoi cell of a new center overlaps on average $M$ preexisting Voronoi cells. Then we can expect that $p^{A}(s)\sim q^{c}(s)\sim s^{\gamma}P(s)$. On the other hand, for large values of $s$, $dP(s)/ds<0$, which means that $p^{A}(s)$ controls the right side of Eq.~(\ref{psevol2d}). Thus, in this limit we can expect that $P(s)\sim \mathrm{exp}\left(-M\,s^{\gamma}/(\gamma\,\tilde{\mu}_{\gamma})\right)$. In our 2D model, the tail of $P(s)$ for large values of $s$ depends on the parameter $\gamma$.

In the opposite limit $s\ll1$, the behavior of $P(s)$ is given by $p^{*}(s)$ because this distribution dominates the right side of Eq.~(\ref{psevol2d}). Then $p^{*}(s)\sim s^{\zeta}$ implies $P(s)\sim s^{\zeta}$. In order to understand the behavior of $p^{*}(s)$ for small values of $s$ we proceed as follows.

In the simplest case a new Voronoi cell is completely defined by just three nearby centers; see Fig.~\ref{cpzab}. Note that the new center shown in Fig.~\ref{cpzab}(b) generates a small Voronoi cell even though it was formed at the expense of the three large Voronoi cells shown in Fig.~\ref{cpzab}(a). If $p_c$ is the probability to have the initial configuration shown in Fig.~\ref{cpzab}(a), then the probability to place a new center as in Fig.~\ref{cpzab}(b) is given by $p_c\,A^{\gamma}$ where $A$ is the area of the target region. Naturally, $A$ scales with the distance between centers as $A\sim r^2$. In the case of noncorrelated centers, $p_c\sim p^{(0)}(r_1)p^{(0)}(r_2-r_1)p^{(0)}(r_3-r_2)\sim c(r_1)c(r_2-r_1)c(r_3-r_2)$ which leads to $p_c\sim r^3$. In the PV case we have $\gamma=1$; thus, $p^{*}(s)\sim s^{2.5}$. Note that this result agrees with Eq.~(\ref{ps2d}). The case of correlated centers is significantly more complicated because $p_c$ cannot be written as an independent product of $p^{(0)}(s)$. In any case, it is clear that $p^{*}(s)$ for small values of $s$ is related with the concentration of centers $c(R)$ through the radial distribution functions which in turn in our model depend on $\delta$. Then, the value of $\zeta$ in $P(s)\sim s^{\zeta}$ for a given value of $\gamma$ increases with $\delta$. However, a general relation between $\delta$ and $\zeta$ remains unknown.

As noted, Eq.~(\ref{psevol2d}) cannot be solved analytically; however, the numerical simulation of this system can be done without major complications \cite{sim2}.

\section{Applications of the NPV patterns}\label{secAp}
\subsection{Gap size distribution of parked cars}
Rawal and Rodgers (RR hereafter) measured the size of the gaps between adjacent parked cars \cite{rawal}. The data (500 gaps) were gathered from four connected streets in London without any side streets or driveways. They found that small and large gaps are unlikely. The effective repulsion between adjacent cars arises because drivers need to leave some space between cars to allow exit maneuvers. RR suggest that large gaps are unlikely because people try avoid the waste of space. However, we believe that it is more reasonable to think that this happens because people usually prefer to park in large gaps where the entry maneuver is the easiest. This implies that large gaps are often destroyed before small ones. RR developed two different models to describe $p^{(0)}(x)$; however, just one of them describes the statistical behavior of the system properly. In their improved model they consider two different factors: people who park anywhere and those who perform an additional maneuver to avoid the waste of space. These assumptions give a good description of the empirical data.

On the other hand, Abul-Magd \cite{abul1} used the Wigner surmise (WS) with $\beta=2$ as an approximate model for $p^{(0)}(x)$. The WS describes the gap distribution of a one-dimensional Coulomb gas at an inverse temperature $\beta$. In this system, particles are free to move around a circle but experience a logarithmic potential interaction. He selected this value of the inverse temperature $\beta$ because in this case the WS describes excellently the statistical behavior of chaotic systems without time-reversal symmetry, as is appropriate for the car-parking problem. The WS does give a good quantitative description of the car-parking problem; however, the physical interpretation of the logarithmic interaction potential among cars is not entirely clear. In Ref.~\cite{seba}, \v{S}eba reported more extensive (over 1200 gaps) empirical data for the gap sizes. He modeled the car parking with a Markov chain where the cars are allowed to get in and get out of the gaps following prescribed probabilistic rules.

\begin{figure}[t]
\begin{center}
\includegraphics[scale=0.3]{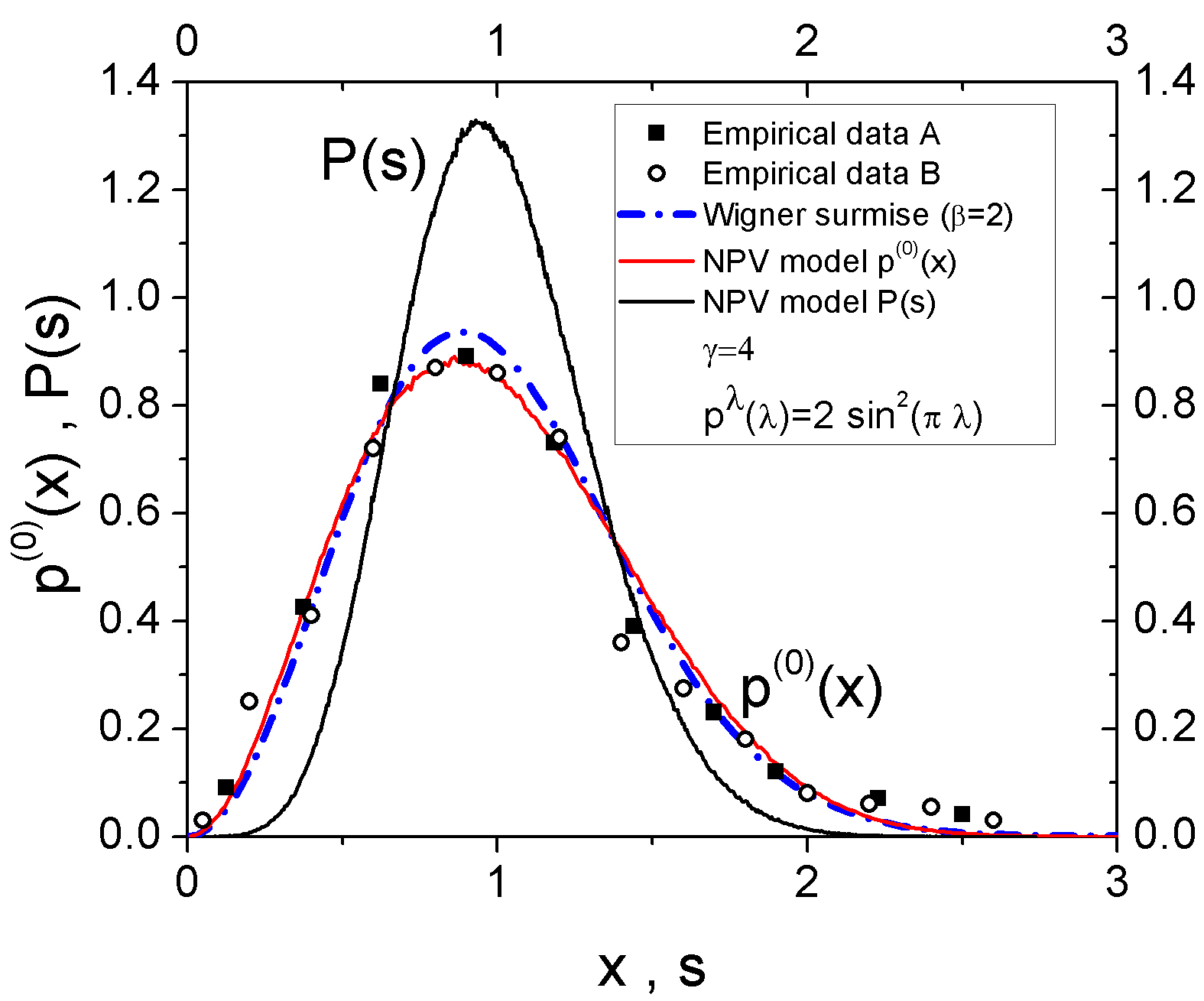}
\end{center}
\caption{(Color online) NPV case with $\gamma=4$ and $p^{\Lambda}(\lambda)=2\,\sin^2(\pi \lambda)$. Here empirical data A and B correspond to the measurements reported in Refs.~\cite{rawal} and \cite{seba}, respectively.}
\label{car}
\end{figure}

From our model to generate one-dimensional Voronoi cells, we can interpret the car-parking problem more simply and intuitively. As in the models mentioned before, our NPV model for car parking contains some simplifying assumptions for an ensemble of drivers such as homogeneity and isotropy. Our goal is to analyze a tractable version for the problem rather than contend with all the subtleties. As mentioned previously, we start with the assumption that people prefer to park in large gaps rather than in small ones. This follows from the fact that the parallel parking is easier in spots where there is space to spare. Thus, it is reasonable to propose that the probability to park in a gap with size $x$ can be modeled by Eq.~(\ref{pa}). Naturally we have to choose the appropriate value of $\gamma$. Once the gap has been selected, the driver has to choose the exact parking place inside the gap. For gaps shorter than two car lengths the most likely place to park would seem to be the middle of the gap. For lack of a better simple ansatz, we extend this assumption to gaps of arbitrary length. Additionally, the driver should avoid parking too close to the cars on the borders of the gap in order to guarantee enough space to leave the gap when necessary. 
This implies $p^{\Lambda}(0)=p^{\Lambda}(1)=0$. Finally, we claim that $p^{\Lambda}(\lambda)=p^{\Lambda}(1-\lambda)$; i.e., the drivers do not have preference to park near to the car on either side of the gap. A simple function which satisfies those properties is $p^{\Lambda}(\lambda)=2\,\sin^2(\pi \lambda)$. From our numerical results we found this functional form for $p^{\Lambda}(\lambda)$ gives a reasonable description of the empirical data when $\gamma=4$.

\begin{figure*}[t]
\begin{center}
$\begin{array}{cc}
\includegraphics[scale=0.25]{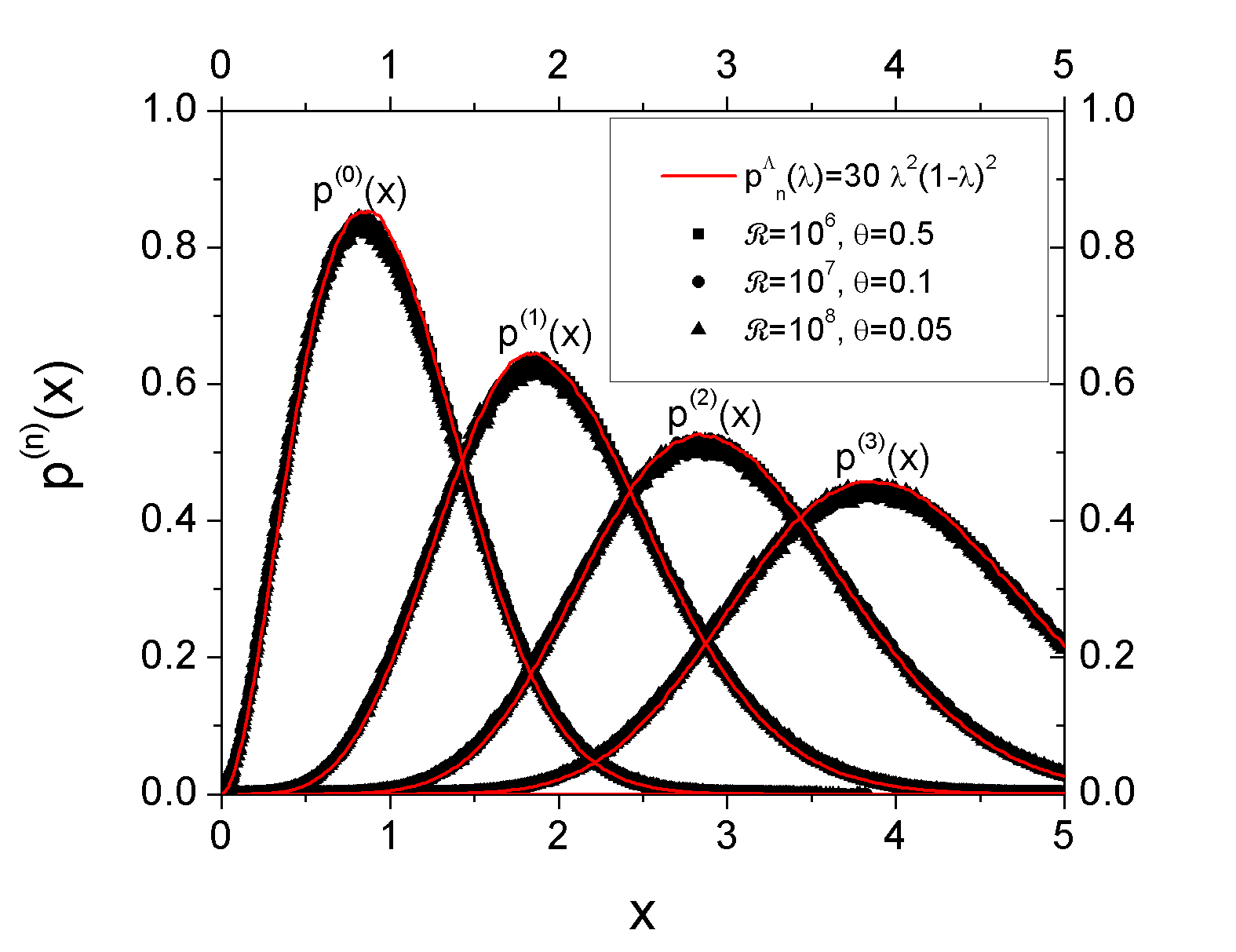}&
\includegraphics[scale=0.25]{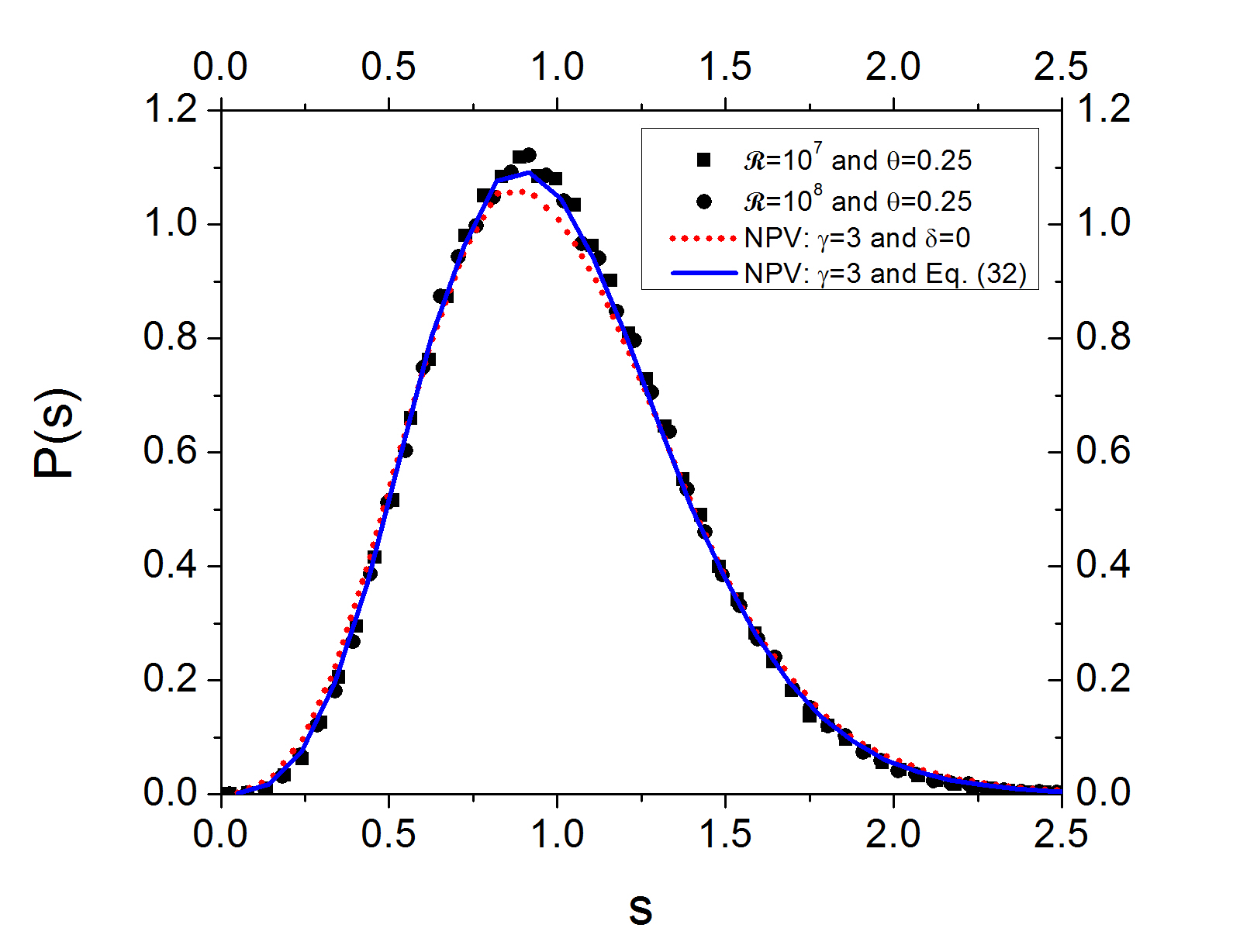}\\
(a) & (b) \\
\end{array}$
\end{center}
\caption{(Color online) The statistical behavior of irreversible nucleation in 1D is shown in (a). The CZ distribution for 2D nucleation is shown in (b). In all cases our results are compared with direct numerical simulations of island nucleation for several values of $\Re$.}
\label{f1}
\end{figure*}

As shown in Fig.~\ref{car}, our model describes excellently the empirical data given in Refs.~\cite{rawal,seba}. Of course, more refined models can be developed, but the most important result is that a suitable selection of $\gamma$ and $p^{\Lambda}(\lambda)$ leads to an excellent description and interpretation of the statistical behavior of the gap size of parked cars. From our previous discussion, it is clear that $p^{(0)}(x)\sim x^{\zeta}$ with $\zeta=2$ for $x\ll1$. The value of $\zeta$ is related to the effective ``repulsion force'' between adjacent cars. In the limit $x\gg1$, $p^{(0)}(x)\sim x^2\,\exp(-x^4/(4\,\mu_{\gamma}))$. Note that $\gamma=4$ represents the strong preference of drivers to park in large gaps rather than in small ones.

In the problem of parked cars, we can interpret the Voronoi size distribution $P(s)$ through the next-nearest spacing distribution $p^{(1)}(x)$ as follows. From its definition, $p^{(1)}(x)$ clearly gives the probability density that a parked car has a spot of length $x$ to perform the exit maneuver. From Eq.~(\ref{CZ1d}), the distribution of the Voronoi cell sizes $s$ is proportional to the distribution of distances $x$ that the drivers have to perform an exit maneuver.

\subsection{Point-island model for epitaxial growth in 1D and 2D}
In the point-island model for epitaxial growth, atoms are deposited onto a substrate where they perform random walks. In the simplest case, when two atoms meet they form a static island. In the same way, the atoms which reach an island are captured and remain attached to it. This case corresponds to irreversible attachment and is usually called ``$i=1$'' in the literature. Another important characteristic of the point-island model is the fact that the islands do not grow laterally. The size of a particular island is given just by the number of atoms which belong to it. This system exhibits a scaling regime in the limit $\Re=F/D\rightarrow\infty$, where $F$ is the deposition rate of atoms and $D$ is their diffusion constant. This model is, of course, a simplification of the real system but it contains most of the relevant physical properties required to describe the processes behind the island formation in epitaxial growth \cite{gonzalez7,blackman,evans5,amar3,amar5,ratsch1,tokar,oliveira}. In fact, the widely used point-island model gives very accurate results in early stages of growth (low coverages) and it is an important theoretical model in our knowledge about epitaxial growth.

In this context, the point islands determine the pattern of Vononoi cells, playing the role of the centers defined in Sec.~\ref{int}. The atoms inside a particular Voronoi cell are usually captured by the center of the cell (island). Because of this, the Voronoi cells are called capture zones (CZs) in the context of epitaxial growth. Naturally, the growth rate of an island is related to the size of its CZ. For more details about these kinds of models see, for examples, Refs.~\cite{gonzalez7,blackman,amar3,evans,amar4,amar5,ratsch1,tokar,oliveira,evansthiel,mulheranrev}.

\subsubsection{One-dimensional case}
Consider now the case of a one-dimensional substrate with irreversible attachment. A suitable choice to describe the spacing and the CZ distributions of the 1D point-island model is \cite{gonzalez7}
\begin{equation}\label{plambda}
p^{\Lambda}(\lambda)=30\,\lambda^2(1-\lambda)^2
\end{equation}
and
\begin{equation}\label{gammad1}
\gamma(x)=\left\{
\begin{tabular}{cc}
$3$,\, & if $x>1.7$\\
$4$,\, & if $x\leq1.7$\\
\end{tabular}
\right.\end{equation}

Blackman and Mulheran \cite{blackman} originally calculated Eq.~(\ref{plambda}) by first obtaining the average density of atoms, $n_1(x,y)$, inside a gap of length $y$ from its expression in the stationary state. From this approximation and assuming that the probability of a new nucleation at $x$ is proportional to $n_1(x,y)^2$, they found Eq.~(\ref{plambda}). On the other hand, Eq.~(\ref{gammad1}) is based on the numerical results reported in Ref.~\cite{gonzalez7}. In Fig.~\ref{f1}(a), the results of this model are shown and compared with the direct numerical simulation of the island nucleation for three different values of $\Re$; the agreement is excellent. This selection of $\gamma(x)$ and $p^{\Lambda}(\lambda)$ reproduce the statistical behavior of the 1D point-island model with irreversible attachment. For additional information see Refs.~\cite{blackman,gonzalez7}.

\begin{figure}[ht]
\begin{center}
\includegraphics[scale=0.3]{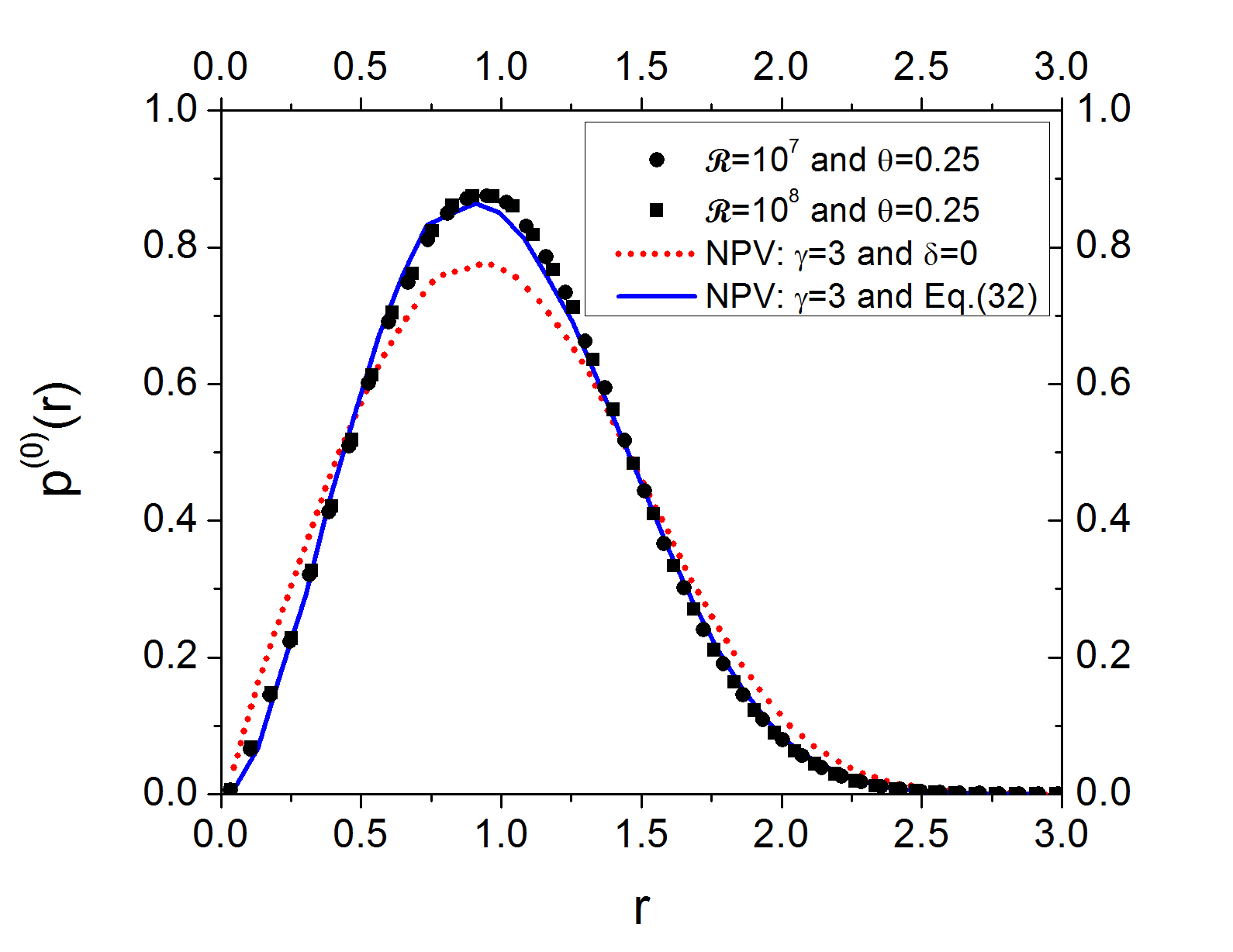}
\end{center}
\caption{(Color online) Behavior of the first radial distribution $p^{(0)}(r)$, which is related to the concentration of islands $c(r)$.}
\label{p02d}
\end{figure}

\subsubsection{2D in circular-cell approximation}
The point-island model with irreversible attachment in two dimensions can be modeled following a similar procedure to that was used previously to described nucleation in one dimension.  To make analytic progress, we follow Refs.~\cite{evans1,evans2,evans3,evans4} by approximating the Voronoi cells as circular. (While Figs.~\ref{pattpv}, \ref{cpzab}, etc., show that typical cells are far from circular, the approximation is better than might be anticipated because the average shape over an ensemble of cells does tend toward circular.) As shown in Eq.~(B2a) of Ref.~\cite{evans1} and Eq.~(9) of Ref.~\cite{evans4}, the isotropic steady-state solution of the appropriate diffusion equation with flux $F$ inside such a circular Voronoi cell with radius $R_c$, for a concentric (non-point) island of radius $R_{isl} < R_c$, gives the following expression for the density of atoms:
\begin{equation}\label{rhoislas}
\rho(R)=\frac{1}{2} \frac{R_c^2}{\Re} \left[\mathrm{ln}\left(\frac{R}{R_{isl}}\right) + \frac{1}{2}\frac{R_{isl}^2}{R_c^2}\left(1-\frac{R^2}{R_{isl}^2}\right)\right],
\end{equation}
where $R$ (assuming $R_{isl} \le R \le R_c$) is the distance from the cell center.  Note that $\rho(R_{isl})=0$ (thence increasing linearly with $R-R_{isl}$ initially) and $\left.d \rho(R)/dR\right|_{R_{c}}=0$. The first condition comes from the density of atoms being zero along the island boundaries. On the other hand, the probability of nucleation is proportional to $\rho^2(R)$, which is maximal along the boundary of the CZ. Hence, $\rho(R)$ is also maximal at $R=R_{c}$, as implied in the second (Neumann) condition.

In this framework [with nucleation $\propto \rho^2(R)$], the probability to have a nucleation event within a particular circular capture zone with radius $R_c$, $P_n(R_s)$, can be written as
\begin{equation}\label{Pn2d}
P_n(S)=\frac{q^{\gamma}(S)}{\hat{P}(S)}\propto \int^{R_c}_{R_{isl}}dR\,2\,\pi\,R\,\rho^2(R)\propto R_c^6.
\end{equation}
Note that $P_n(S)$ is proportional to the third power of the scaled area of the cell, i.e., $\gamma=3$. This is, of course, a strong approximation. It is well known that the radius of a capture zone fluctuates significantly around its mean value because the centers are usually not at the geometric center of the cells. Approximating the shape of a CZ by a circle neglects those radial fluctuations. Furthermore, the density of atoms inside a CZ depends on the position of neighboring centers, which is not taken into account by Eq.~(\ref{rhoislas}). This also implies that the isotropy assumption used to write Eq.~(\ref{palfa}) is poor. From Eq.~(\ref{Pn2d}) it is clear that the probability of nucleation increases with the distance from the center and reaches its maximum along the boundary of the capture zone. There are many ways to select the place of nucleation inside a particular capture zone \cite{evans2,evans3,evans4}. However, in order to keep our model as simple as possible, we assign the same probability to all points inside a particular CZ, regardless of their distance from the center; i.e., $\alpha=0$. While this is a crude approximation, we can see in Fig.~\ref{f1} that it is adequate to describe $P(s)$. However, previous simplifications have important effects on the radial distributions. Figure~\ref{p02d} shows the behavior of $p^{(0)}(r)$. Not unexpectedly, our simplified model does not describe $p^{(0)}(r)$ appropriately; i.e., it is not a good approximation for the island density $c(R)$.
From Eq.~(\ref{rhoislas}) it is clear that the concentration of islands in the limit $R\ll1$ is given by $c(r)\sim R^2$, which implies $\delta=1$.

In order to improve our model, we must take into account the fact that $\rho(R)$ vanishes along the boundaries of the islands; i.e., $q^{r}(0,s)=0$. Perhaps the simplest way to accomplish this goal for the point-island model is to propose
\begin{equation}\label{pcisd2d}
q^{r}({\rm {\bf R}},s)\sim\left\{
\begin{tabular}{cc}
$R$,\, & if $0\leq R\leq\kappa \,R_c$\\
$\kappa\,R_c$,\, & \,\,\,\,if $\kappa\,R_c< R\leq R_c$\\
\end{tabular}
\right.\end{equation}
where $0< \kappa<1$ is a constant, and $R_c=(S/\pi)^{1/2}$ is the average radius of the CZ. From our numerical experimentation we estimate $\kappa=0.3$. In this way, the nucleation probability inside a capture zone grows linearly with $R$ for points near the island, while it becomes constant for points far away.


Figures~\ref{f1}(b) and \ref{p02d} also include the results of this last model. The description of $P(s)$ is again excellent, but now the nearest-neighbor radial distribution is also well fitted. Thus, in our model for the island nucleation in 2D, the repulsion between centers given by $\alpha$ seemingly has a great impact on $p^{(0)}(r)$ but is not crucial to determine $P(s)$.

\begin{figure*}[htp]
\begin{center}
$\begin{array}{cc}
\includegraphics[scale=0.3]{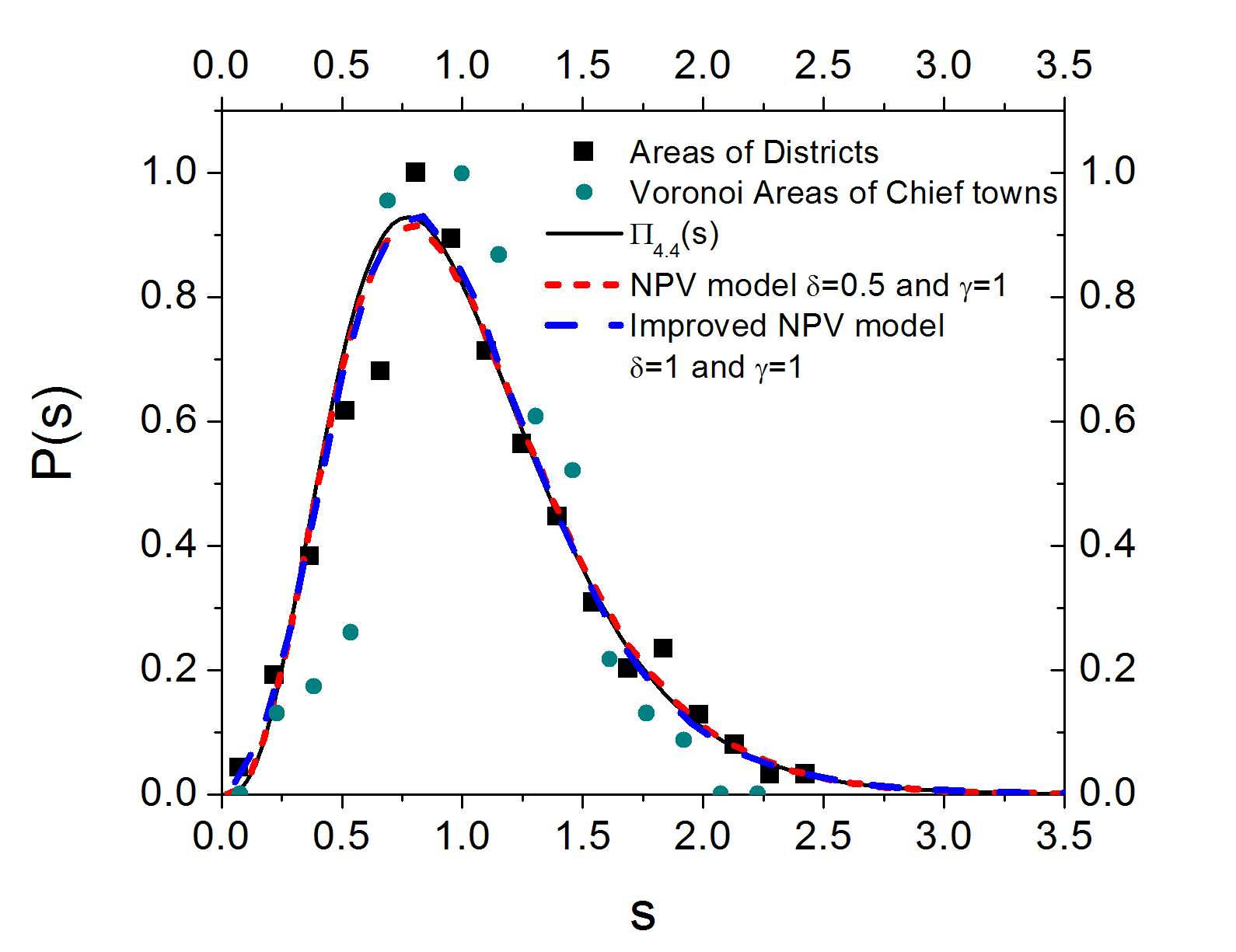}&
\includegraphics[scale=0.23]{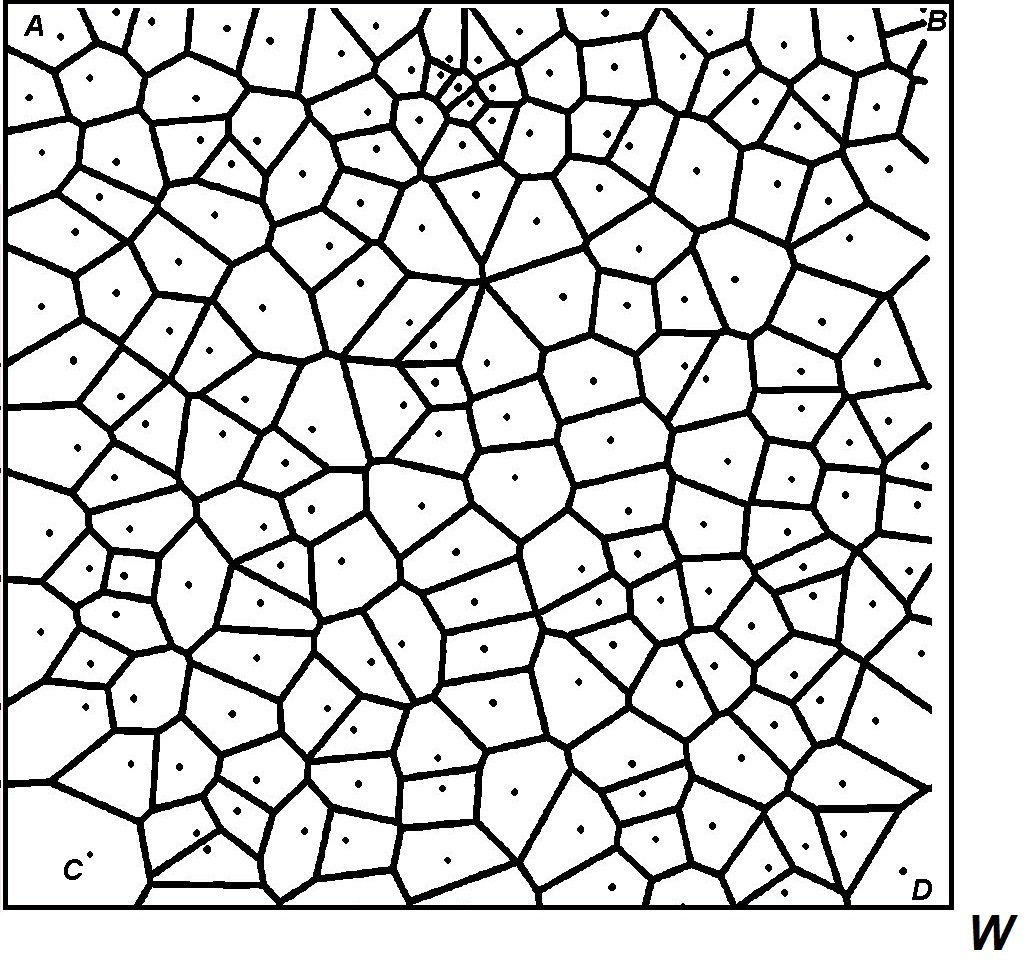}\\
(a) & (b) \\
\end{array}$
\end{center}
\caption{(Color online) (a) Size distribution of the 322 districts of mainland France (excluding Paris). (b) NPV pattern generated by the 188 chief towns in the districts in a rectangle $W$ in France, as discussed in Ref.~\cite{caer}.  The points A, B, C, and D represent Saint-L\^o, Thionville, Mont-de-Marsan, and Apt, respectively. (The cluster of points in the upper part of the figure represents Paris and the nearby chief towns.) The area distribution of these Voronoi cells is included in panel (a) and is rather similar to the distribution of actual districts for 1/2 $< s <$ 2.  The fits are done iteratively.  The improved NPV model [cf.\ Eqs.~(\ref{pgammar}) and (\ref{palphar})] takes into account the finite area of the chief towns by assuming a core radius that is 2/5 of the mean radius.}
\label{slad}
\end{figure*}

\begin{figure*}[ht]
\begin{center}
$\begin{array}{cc}
\includegraphics[scale=0.3]{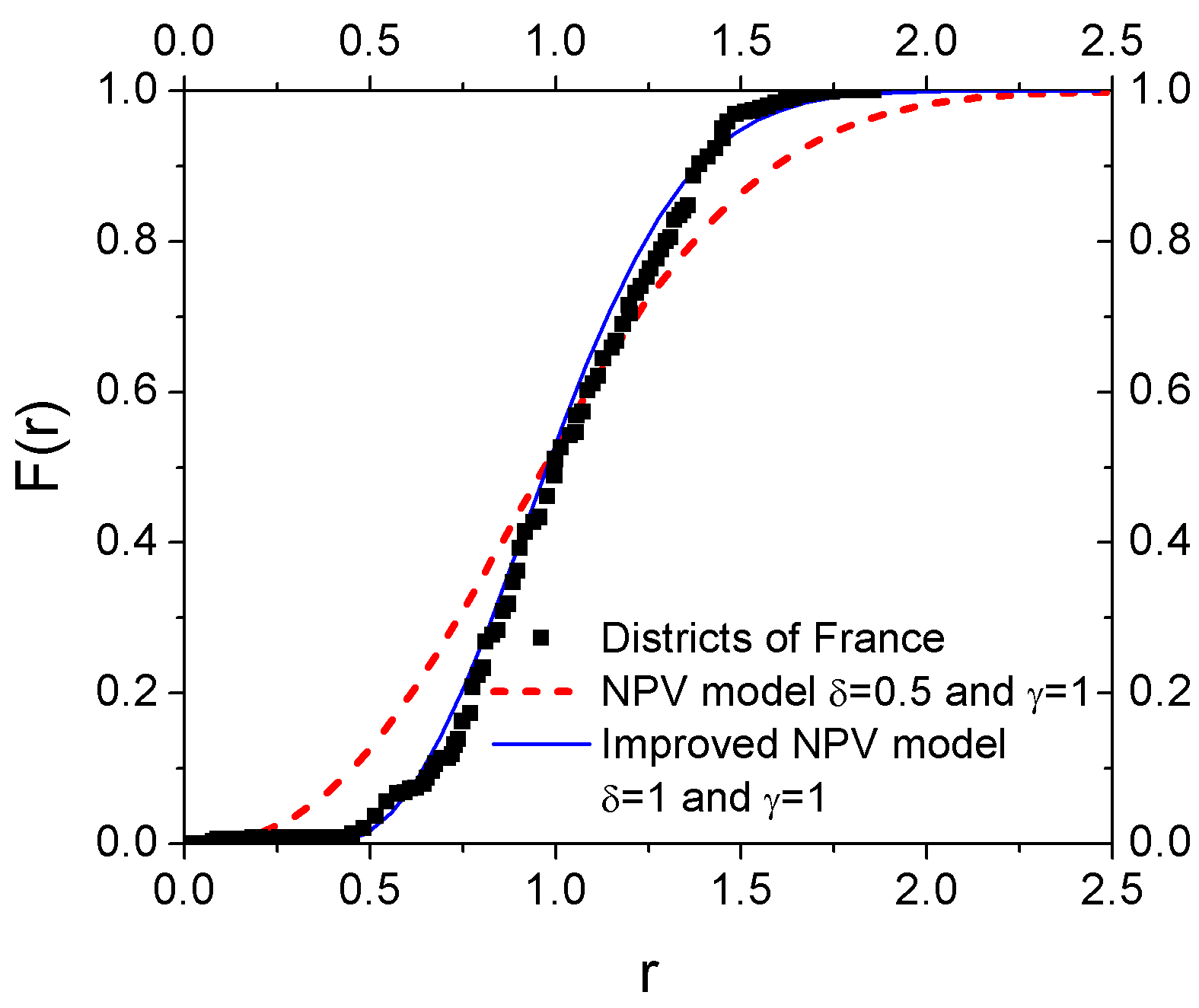}&
\includegraphics[scale=0.3]{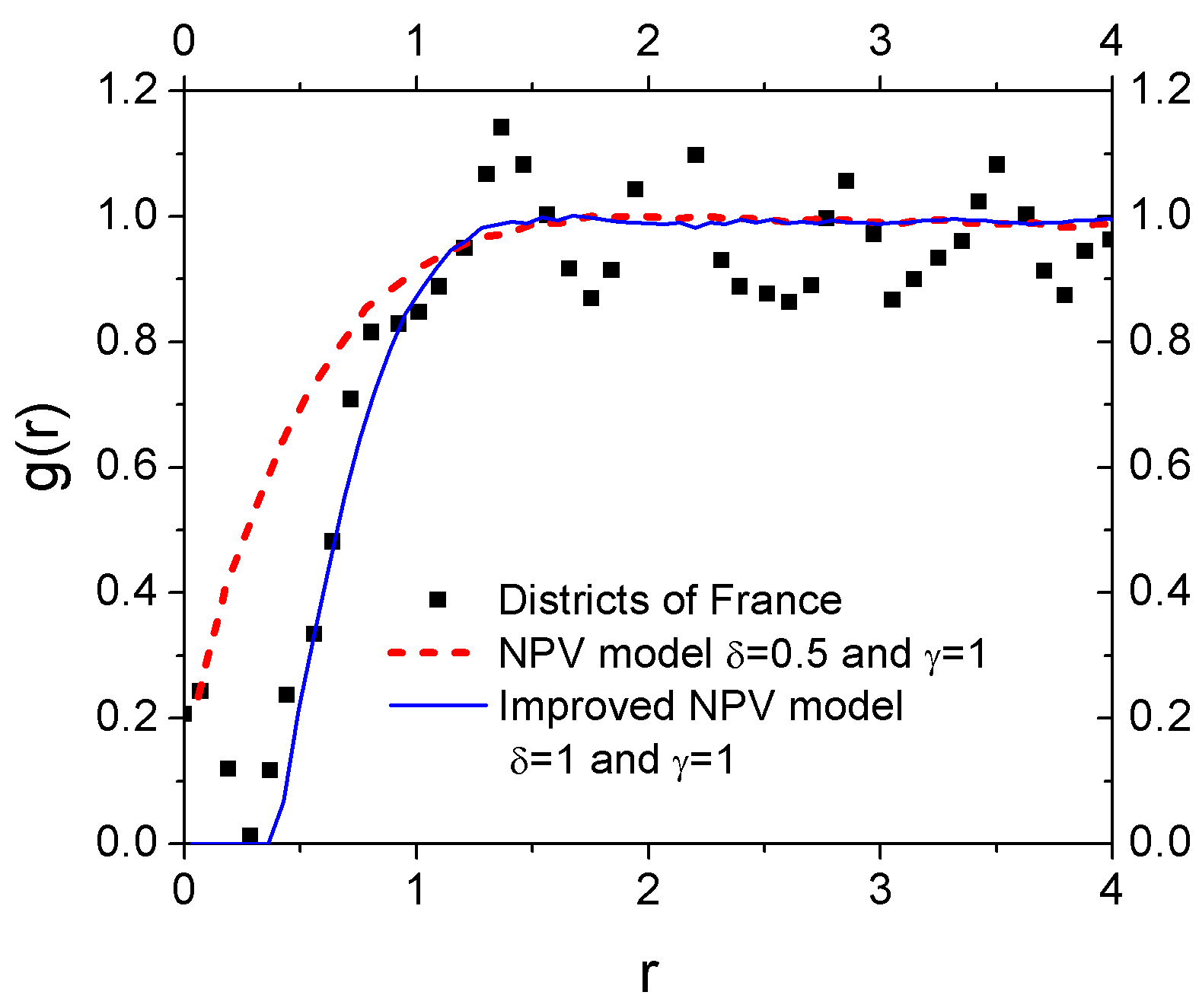}\\
\end{array}$
\end{center}
\caption{(Color online) (a) Cumulative distribution $F(r)$ and (b) pair correlation function $g(r)$, based on the set of points depicted in Fig.~\ref{slad}(b), using the point-island NPV model and the improved NPV model incorporating the finite size of chief towns.}
\label{slad1}
\end{figure*}

\subsection{Size distribution of second-level administrative divisions}
The polygons formed by county boundaries in the size-division model resemble Voronoi cells \cite{newberry,caer}. Inspired by this fact, we explore the possibility to use our NPV model to study the formation of second-level administrative divisions (SLADs), such as counties in the USA or (non-urban) districts ({\it arrondissements}) in France. The formation of SLADs depends on many political, cultural, ecological, and geographical factors \cite{caer,rajesh}. These factors are in general difficult to include in a mathematical model. However, our model allows us to give a simple interpretation of the SLAD formation in terms of $q^{c}(s)$ and $q^{r}({\rm {\bf r}},s)$.

We focus here on the results reported by Le Ca\"er and Delannay (LCD) for the SLADs in France \cite{caer}. The departments
({\it d\'epartements}) comprise the first administrative division in France.  In mainland France there are 94 departments. Each department is divided into several districts; there are a total of 322 districts in mainland France.  Each district has a chief town, mostly with the same name as the district.

\begin{figure*}[t]
\includegraphics[scale=0.16]{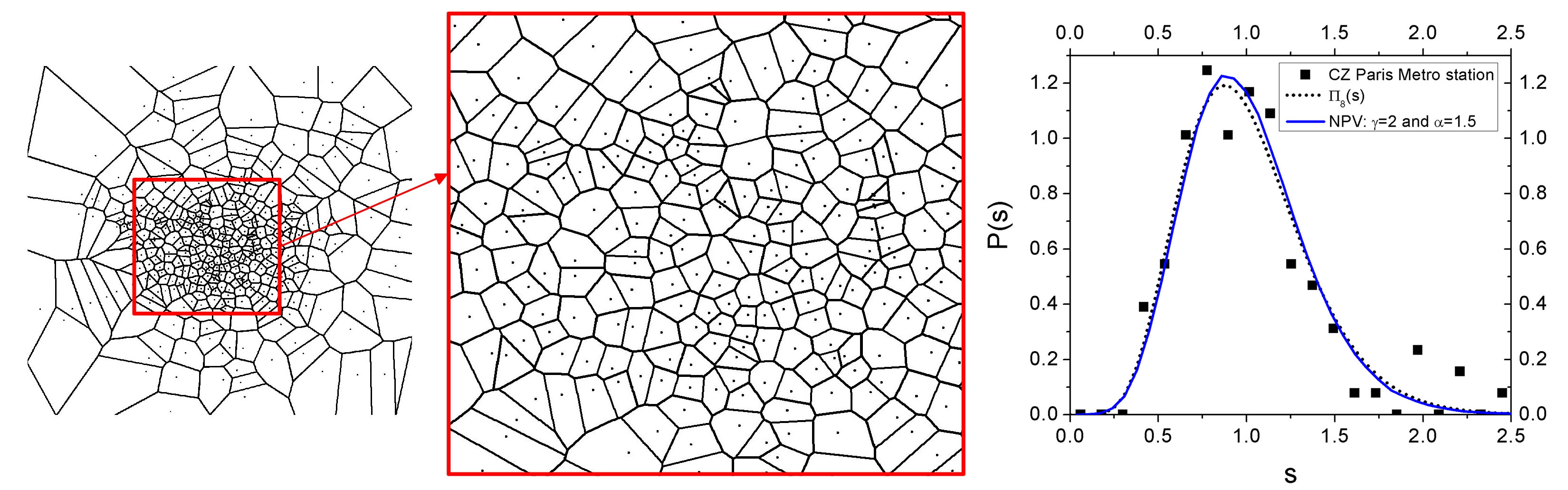}\\
(a) \hspace{7cm}  (b)
\caption{(Color online) (a) Voronoi cells (CZ) for the Paris M\'etro stations.  The complete network of stations on the left is enlarged to show the stations and associated cells near the heart of the city. (b) Capture-zone distribution of the Paris M\'etro stations.}
\label{paris}
\end{figure*}

LCD compared the area distribution of the French SLADs \cite{caer,wikiarron} to the one-parameter gamma distribution $\Pi_{\alpha}(s)$ with $\alpha\approx4.4$ \cite{fgama}. As seen in Fig.~\ref{slad}(a), the agreement between $\Pi_{4.4}(s)$ and the data is satisfactory. Note that $\alpha=4.4$ is larger than the value used to describe the size distribution of PV cells ($\alpha=2.5$). In our model, this implies $\delta>0$. It follows that the SLAD formation process in the rectangle $W$ was not a completely random process. In fact, taking $\delta=0.5$ and $\gamma=1$ we found good agreement between the data and the results of our model. The value of $\gamma$ was chosen to have an exponential tail for $P(s)$ such as in Eq.~(\ref{ps2d}).

In particular LCD focused on the pattern formed by the chief towns of the districts which are located within a 544 km $\times$ 636 km rectangle whose corners are close to the towns Saint-L\^o, Thionville, Apt, and Mont-de-Marsan.  The average distance between the enclosed 188 chief towns is 29.6 km. As shown in Fig.~\ref{slad}(b), the pattern formed by the chief towns looks similar to that in Fig.~\ref{pattpv}. However, as noted by LCD, the formation of these SLADs was not a PV process.

The parameters $\delta=0.5$ and $\gamma=1$ do not describe the cumulative nearest-neighbor distribution,
\begin{equation}
F(r)=\int^{r}_0\,d\xi\,p^{(0)}(\xi),
\end{equation}
nor the pair correlation function,
\begin{equation}
g(r)=\frac{1}{\rho \,N} \left.\sum\right._{i\ne j}^{N} \langle \delta({\rm {\bf r}}+
{\rm {\bf r}}_j-{\rm {\bf r}}_i)\rangle
\end{equation}
as seen in Fig.~\ref{slad1}. Increasing the value of $\delta$ improves the estimation of $F(r)$ and $g(r)$ but it leads to a poor description of $P(s)$. This means that a model like ours is insufficient for this kind of system.

We attribute this discrepancy to the assumption of point islands. One must account for the actual areas of cities, which produces an effective  short-range repulsion. The simplest way to incorporate this feature into our model is to introduce an excluded area around each city center in the form of a hard-core radius $r_{core}$. In particular, we modify Eqs.~(\ref{pgamma}) and (\ref{palfa}) as follows:
\begin{equation}\label{pgammar}
q^{c}(s)=\frac{\left(s-\pi\,r_{core}^2\right)^{\gamma}}{\tilde{\mu}_{\gamma}} P(s)\Theta(s-\pi\,r_{core}^2),
\end{equation}
and
\begin{equation}\label{palphar}
q^{r}({\rm {\bf r}},s)\sim \left(r-r_{core}\right)^{\delta}\Theta(r-r_{core}),
\end{equation}
where $\Theta(\xi)$ is the unit step function. When $r_{core}=0$, we recover the original model. From numerical simulation of this improved model, we found excellent agreement with the data by using $r_{core}/\left\langle r \right\rangle\approx0.4$, $\delta=2$, and $\gamma=2$ (see Figs. \ref{slad} and \ref{slad1}). This improved model still describes adequately $P(s)$, but now the fits involving $p^{(0)}(r)$ and $g(r)$ are significantly better. Note that this model does not describe the cluster given by Paris and the nearby towns. Clearly, the population density is highest in Paris and its surroundings. Consequently, it is reasonable to expect the formation of clusters of towns in this region of France, which is inconsistent with our ansatz of an excluded area.

Our analysis has been applied to SLADs in some 20 countries, with results generally consistent with the above
analysis \cite{rajesh2}. As we report in Ref.~\cite{rajesh2}, however, there are some subtleties and a rich range of nuances, e.g., regional differences in the distributions of the areas of counties in the USA.

\subsection{Capture zones of Paris M\'etro station}
In analogy with the island nucleation, we can define the Voronoi cells or CZ for M\'etro stations as follows. Each M\'etro station represents a center. The M\'etro stations are in competition for passengers in the same way that the islands compete for atoms. If we suppose that all M\'etro stations are accessible from anywhere, then most of the passengers within a particular Voronoi cell will be ``captured'' by the center of this cell. Of course, this is an oversimplification. The passengers not only selected a M\'etro station because of its proximity. They also take into account ease of access to it (parking, bus routes, commuting possibilities, road conditions, etc.).

As with the SLADs, in order to have a good set of data to apply our model, we have to select a city with a near-uniform geographical profile and with a large number of M\'etro stations. These conditions are approximately satisfied by the capital of France, Paris. Figure~\ref{paris}(a) shows a scale diagram of the M\'etro stations of Paris (RER stops are also included). Clearly the density of centers (M\'etro stations) is not constant. Naturally, there are more M\'etro stations near the center of the city, where the population density is highest. Because of this, the largest Voronoi cells are near the outskirts of the city. However, we can expect that our NPV model works well if we just consider stations near the city center, where the density of M\'etro stations does not change dramatically, as seen in the enlargement in the figure.

For economic reasons it is unlikely that two or more M\'etro stations are very close together. We then expect $\alpha>0$ because there is an effective ``repulsion force'' between stations. In Fig.~\ref{paris}(b), we compare the empirical CZ distribution of the in-town M\'etro system with our NPV model. Good agreement is found with $\gamma=2$ and $\alpha=1.5$; the gamma function $\Pi_{8}(s)$ is also included in the comparison presented in Fig.~\ref{paris}(b). Apparently the ``repulsion force'' between M\'etro stations is bigger than that for island nucleation in 2D.

\section{Conclusions}\label{secCon}
Our proposed NPV model can be used to describe and interpret many different systems in terms of two independent distribution functions. In 1D, $p^{g}(x)$ gives the probability density to put a new center inside a gap with a scaled size $x$. This distribution is given explicitly by Eq.~(\ref{pa}) and is controlled by the parameter $\gamma$. This parameter determines the behavior of the gap size distribution for large values of $x$. In fact, for $x\gg1$, $p^{(0)}(x)\approx A\,x^{-2}\mathrm{exp}\left(-x^{\gamma}/(\gamma\,\mu_{\gamma})\right)$, independent of the kernel $p^{\Lambda}(\lambda)$. Additionally, $\gamma$ modulates the size dependence of the destruction of gaps. For $\gamma=0$, the probability to put a new center within a gap is the same for all gaps, regardless of their size. The larger $\gamma$ is, the greater is the probability of destruction of large gaps. For the car-parking problem we use $\gamma=4$, which reflects the preference of drivers to park in large gaps rather than small ones. For island nucleation in 1D it is necessary to take into account that $\gamma$ is a function of the scaled size $s$.

In 1D, the behavior of $p^{(0)}(s)$ in the limit $s\ll1$ is completely determined by the fragmentation kernel $p^{\Lambda}(\lambda)$. For the kernels considered here, we always have the generic behavior $p^{\Lambda}(\lambda)\sim \lambda^{\zeta}$ for $\lambda \ll1$. The parameter $\zeta$ controls the effective repulsion force between centers. For island nucleation in 1D and the car-parking problem, we found $\zeta=2$. For the first system the value of $\zeta$ is fully determined by the density of atoms inside the gap in the aggregation regime. In the car-parking problem, $\zeta$ reflects the need of the drivers of allow space between cars to perform an exit maneuver.

In 2D, the probability density $q^{c}(s)$ to put a new center inside a Voronoi cell is also controlled by the parameter $\gamma$. In the case of the SLADs, $\gamma=1$ gives a good fit of the empirical data. Because of this, the $P(s)$ of many SLADs can be approximated by using the single-parameter gamma distribution $\Pi_{\alpha}(s)$ \cite{rajesh}. In the case of Paris M\'etro stations we used $\gamma=2$.

The position {\bf r} of a new center inside a particular Voronoi cell in 2D is determined in our model through $p^{\delta}({\rm {\bf r}},s)$. For the sake of simplicity, we consider only isotropic cases. This probability is closely related to the concentration of centers $c(R)$. For example, in the case of island nucleation in two dimensions, a good description of $p^{(0)}(r)$ requires taking into account that the density of atoms vanishes along the island boundaries, even though the CZ distribution can be well described without taking into account this fact. This suggests that many different fragmentation models can be used to describe the CZ distribution of islands in epitaxial growth. However, just a few of them fully describe the statistical behavior of the system in a proper way. An additional example is given by the SLADs in France. Two different sets of parameters describe $P(s)$ properly but just one of them gives a good fit for $F(r)$ and $g(r)$.

We defined the CZs of M\'etro stations; an extension to other systems defined by gas stations, public schools, coffeehouses, post offices, etc., is straightforward. In epitaxial growth, the CZ of an island is related to the islands rate of capturing atoms. It is reasonable to expect that the number of passengers entering a M\'etro station or the influx of customers patronizing a retailer is intimately related to the size of its CZ.

The model presented here allows us to describe quantitatively and qualitatively many systems based on simple assumptions about them. For example, in the car parking problem we based our model on some assumptions of the driver preferences related with parallel parking. Our ansatz leads to a reasonable description of the gap size distribution. For the description of the CZ distributions in the point-island model, we based our simulation on an estimate of the density of atoms (besides other observations), which comes from the direct numerical simulation of this system. Nevertheless, there are systems where the NPV patterns are determined by factors difficult to establish as a mathematical expression. In the case of the Paris M\'etro stations, it is clear that there is an ``effective repulsion'' between stations because it is not economically viable to put two or more stations too close. In our model this implies $\delta>0$. However, there are other political, historical, and geographical factors which also affect the CZ formation. Something similar happens in the case of the SLADs.

Despite its implicit simplifications (such as homogeneity and isotropy), our model proves to be a powerful tool to describe several complex systems which are defined through an array of points.

\begin{acknowledgments}
This work was supported by the NSF-MRSEC at the University of Maryland, Grant No.\ DMR 05-20471, and a DOE-BES-CMCSN grant, with ancillary support from the Center for Nanophysics and Advanced Materials (CNAM). The authors thank Alberto Pimpinelli and Rajesh Sathiyanarayanan for valuable comments and observations.
\end{acknowledgments}

\appendix*
\section*{APPENDIX: RELATION BETWEEN $\hat{p}^{(0)}(R)$ and $c(R)$}\label{app}
For an arbitrary isotropic concentration in a circle of radius $R_0$, the expected number of particles inside a concentric disk with radius $R \le R_0$ is $\int^{R}_{0}dr \, 2\pi r\,c(r)$. The probability that this disk contains no particles is
\begin{equation}
E(R)=\left(1-\frac{\int^{R}_{0}dr \,r\,c(r)}{\int^{R_0}_{0}dr \,r\,c(r)}\right)^{\int^{R_0}_{0}dr \, 2\pi r\,c(r)},
\tag{A1}
\end{equation}
when $R\ll R_0$. This last statement leads to
\begin{equation}
\frac{\int^{R}_{0}dr \,r\,c(r)}{\int^{R_0}_{0}dr \,r\,c(r)}\ll1.
\tag{A2}
\end{equation}
Then, we have (for $R\ll R_0$)
\begin{equation}\label{app1}
E(R)=e^{-\int^{R}_{0}dr \, 2\pi r \, c(r)}.
\tag{A3}
\end{equation}

By definition, $E(R)$ is the probability to have an empty disk with radius $R$; then  $E(R)-E(R+dR)$ is the probability of have an empty disk with some particles between $R$ and $R+dR$; i.e., $\hat{p}^{(0)}(R)dR$. We conclude that
\begin{equation}\label{app2}
\hat{p}^{(0)}(R)=-\frac{dE(R)}{dR}.
\tag{A4}
\end{equation}
Equation~(\ref{pcr}) can be obtained from Eqs.~(\ref{app1}) and (\ref{app2}).


\begin{thebibliography}{}
\bibitem{sim1} A lattice with $N$ sites is taken and an array of points on the lattice is generated. Each new point is introduced via weighted random number generation. First, a pair of existing points is selected between which to introduce the new one. The selection is weighted by the $\gamma^{th}$ power of the separation of the points. Then the actual position in the gap for insertion, $x$, is selected with weighting given by the function $p^{\lambda}_{n}(x/y)$. We use $L=10^4$ and gather data for $100$ particles over $5\times10^4$ realizations.
\bibitem{fgama} The gamma distribution was also used in Refs.~\cite{rajesh,rajesh2} to describe the area distribution for the original Thirteen Colonies that evolved into the USA.
\end{thebibliography}
\end{document}